\documentclass[10pt,twocolumn]{article}

\usepackage[T1]{fontenc}
\usepackage[utf8]{inputenc}
\usepackage{newtxtext,newtxmath}
\usepackage[letterpaper,margin=0.85in,columnsep=0.3in]{geometry}
\usepackage{microtype}
\usepackage{booktabs}
\usepackage{array}
\usepackage{amsmath}
\usepackage{graphicx}
\usepackage{xcolor}
\usepackage{enumitem}
\usepackage{tikz}
\usetikzlibrary{arrows.meta,positioning,calc}
\usepackage[numbers,sort&compress]{natbib}

\usepackage{xurl}
\usepackage{flafter}
\usepackage{float}
\usepackage[hidelinks]{hyperref}
\usepackage[capitalise]{cleveref}

\setlist{nosep,leftmargin=*}
\newcommand{\sys}{NoCherry}

\newcommand{\hc}{h_{\mathsf{c}}}
\newcommand{\R}{\rho}
\newcommand{\asm}[1]{\hyperref[asm:#1]{A#1}}

\title{Verifiable Random Sampling of Private Data Without a Trusted Single Party}

\author{Muhammed Karamuk, Miko{\l}aj Glinka, Ruben Wolff\\Forest AI}

\date{}

\begin{document}
\maketitle

\begin{abstract}
\noindent\textbf{Problem:}
When a party supports claims about a private dataset by disclosing a
sample, auditors cannot tell whether it was drawn at random or
``cherry-picked'' to hide flaws in the rest of the data.

\smallskip\noindent\textbf{Solution:}
The data holder commits to the hashes of every item in its dataset and
registers the commitment with independent servers. Selection then waits
on a decentralized randomness beacon: a network that jointly emits a
fresh random value every few seconds, outside the control of the data
holder and the auditors. The value emitted two beacon rounds after the
commitment picks the sample. Each server returns a signed receipt, so
anyone can later check that the revealed sample is exactly the one that
value prescribes. Selection from the committed set stays unbiased as long as the
beacon's threshold holds and one required server is honest, and a
pre-agreed name, checked with the servers, stops repeated or abandoned draws. We call this protocol
\sys{}.

\smallskip\noindent\textbf{Applications:}
Any audit that spot-checks a private corpus: private AI benchmarks and
held-out test sets, dataset quality claims, code and security audits,
test-fairness review, vulnerability handling, and compliance sampling.
Auditors need not trust the data holder to hand over a representative
sample of the committed data, since the protocol proves it, and data
holders need not expose their entire dataset.
\end{abstract}

\section{Introduction}
\label{sec:intro}

Private AI benchmarks pose an audit problem: publishing test items risks
training-data contamination, but keeping them secret prevents outsiders
from inspecting their quality. A maintainer must therefore support claims
about data it cannot fully disclose. Similar constraints arise in audits
of proprietary code, test fairness, and vulnerability handling.

The usual answer is to disclose a random sample. Yet if the data holder
draws the sample, nobody else saw the coin, and it has an incentive to
cherry-pick \citep{ioannidis2005,head2015,cordero2006}. If a single
external auditor draws the sample or timestamps the dataset, outsiders
must trust that single entity not to collude or err. And because digital
data can be changed or fabricated after the holder learns what will be
inspected, the dataset must be cryptographically fixed before the random
selection is known.

\paragraph{Decentralizing trust.}
\sys{} eliminates the single trusted party by splitting trust across two
decentralized components: a threshold randomness beacon (so no single party
controls or predicts the draw) and a set of independent registration
servers (so no single operator is trusted to attest when the commitment
existed). The data holder, called the \emph{committer}, commits to its
items and registers the commitment in parallel with $\ell$ independent
servers. A future beacon round that appears only after registration
deterministically selects which items the committer must reveal. Each
server returns a signed receipt; anyone holding the receipts and revealed
items can verify the entire chain offline. Two attacks matter most:
backdating a commitment after the draw is known, and registering several
commitments under one name to show only the best. Both are prevented as
long as \emph{one} of the $\ell$ required servers is honest.

Decentralizing registration across servers that do not communicate
introduces a coordination problem: network jitter near a beacon tick can
cause different servers to assign the same commitment to different beacon
rounds and thus different random samples. \sys{} resolves this with a
signed \emph{anchor round} that fixes the selection round across all
accepting servers without inter-server consensus.

\sys{} builds on established commit-then-sample methods
\citep{eastlake2004rfc3797,nistbeaconapps,kroll2017accountable} and
realizes the audit mechanism sketched in the PeerBench position paper
\citep{cheng2025peerbench}. Our contributions are:
\begin{enumerate}
\item \textbf{Protocol.} Signed commitments and self-contained receipts
  support offline verification across independent servers. A signed
  \emph{anchor round} guarantees round agreement across all accepting
  servers without communication (\cref{prop:agreement}).
\item \textbf{Analysis.} We state security goals and trust assumptions,
  derive detection probabilities, and analyze repeated commitments.
  \emph{Named commitments} restrict repeated draws under an agreed name;
  their protection depends on server policy and submitter identity
  (\cref{sec:grinding}).
\item \textbf{Implementation and evaluation.} We check independent
  implementations, measure verification costs, and quantify round
  disagreement and the availability cost of anchor freshness
  (\cref{sec:lessons}).
\end{enumerate}
The guarantees concern the committed set; establishing that it is the
right population and ensuring disclosure require external procedures.

\begin{figure*}[t]
\centering
\resizebox{\textwidth}{!}{%
\begin{tikzpicture}[
  font=\footnotesize,
  >=Stealth,
  node box/.style={draw=black!75, rounded corners=3pt, fill=white, align=center, inner sep=4pt},
  group box/.style={draw=black!55, dashed, rounded corners=4pt, fill=black!2, inner sep=0pt},
  arrow/.style={->, thick, draw=black!80},
  beacon arrow/.style={->, thick, densely dashed, draw=black!75}
]

\draw[group box] (-1.75, -2.50) rectangle (1.75, 2.50);
\node[font=\footnotesize\bfseries] at (0, 2.18) {Committer $C$ (Data Holder)};

\node[node box, fill=black!6, minimum width=3.1cm, minimum height=0.90cm] (priv_data) at (0, 1.22) {
  \textbf{Private Dataset} $D$\\[-1pt]
  {\scriptsize Items $d_1, d_2, \dots, d_n$ \emph{(confidential)}}
};

\node[node box, minimum width=3.1cm, minimum height=1.35cm] (commit_op) at (0, -0.48) {
  \textbf{Salted Commitment}\\[-1pt]
  {\scriptsize $c_i = \mathrm{H}(\mathit{tag} \,\|\, s_i \,\|\, h_i)$}\\[-1pt]
  {\scriptsize Sign $(c_1\!\dots\!c_n,\, a,\, K,\, X,\, N)$}\\[-1pt]
  {\scriptsize Commitment hash $\hc$}
};

\draw[->, draw=black!60, thick] (priv_data.south) -- (commit_op.north);

\node[font=\scriptsize, align=center, text=black!80] at (0, -1.92) {
  Discloses \textbf{only} beacon-selected\\[-1pt]
  items $\{d_i, s_i\}_{i \in \mathrm{Sel}}$ at step 4
};

\draw[group box] (4.65, -2.50) rectangle (9.15, 2.50);
\node[font=\footnotesize\bfseries] at (6.9, 2.18) {Independent Registration Servers};
\node[font=\scriptsize\itshape, text=black!75] at (6.9, 1.80) {(1-of-$\ell$ honest assumption; no inter-server communication)};

\node[node box, minimum width=4.15cm, minimum height=1.0cm] (s1) at (6.9, 0.98) {
  \textbf{Server $S_1$} {\scriptsize (Operator 1)}\\[-1pt]
  {\scriptsize Verify $t_{\mathrm{reg}} + \delta_{\mathrm{margin}} < T(a{+}2)$, name $(K,X,N)$;}\\[-1pt]
  {\scriptsize compute $\mathrm{Sel}_\mu(c,\R_{a+2})$, sign $\mathsf{rcpt}_1$}
};

\node[node box, minimum width=4.15cm, minimum height=1.0cm] (s2) at (6.9, -0.22) {
  \textbf{Server $S_2$} {\scriptsize (Operator 2)}\\[-1pt]
  {\scriptsize Verify $t_{\mathrm{reg}} + \delta_{\mathrm{margin}} < T(a{+}2)$, name $(K,X,N)$;}\\[-1pt]
  {\scriptsize compute $\mathrm{Sel}_\mu(c,\R_{a+2})$, sign $\mathsf{rcpt}_2$}
};

\node[font=\footnotesize\bfseries] at (6.9, -0.98) {$\vdots$};

\node[node box, minimum width=4.15cm, minimum height=1.0cm] (sl) at (6.9, -1.82) {
  \textbf{Server $S_\ell$} {\scriptsize (Operator $\ell$)}\\[-1pt]
  {\scriptsize Verify $t_{\mathrm{reg}} + \delta_{\mathrm{margin}} < T(a{+}2)$, name $(K,X,N)$;}\\[-1pt]
  {\scriptsize compute $\mathrm{Sel}_\mu(c,\R_{a+2})$, sign $\mathsf{rcpt}_\ell$}
};

\draw[group box] (11.95, 0.22) rectangle (16.85, 2.50);
\node[font=\footnotesize\bfseries] at (14.4, 2.18) {Decentralized Randomness Beacon $B$};
\node[font=\scriptsize\itshape, text=black!75] at (14.4, 1.80) {(Threshold-BLS nodes, e.g., League of Entropy / drand)};

\node[node box, minimum width=1.15cm, minimum height=0.48cm, font=\scriptsize] (b1) at (12.75, 1.15) {Node $b_1$};
\node[node box, minimum width=1.15cm, minimum height=0.48cm, font=\scriptsize] (b2) at (14.15, 1.15) {Node $b_2$};
\node[font=\scriptsize\bfseries] at (15.05, 1.15) {$\dots$};
\node[node box, minimum width=1.15cm, minimum height=0.48cm, font=\scriptsize] (bm) at (16.05, 1.15) {Node $b_m$};

\node[font=\scriptsize, align=center] at (14.4, 0.54) {
  At $T(a{+}2)$: emit unique, unpredictable $\R_{a+2}$
};

\draw[group box] (11.95, -2.50) rectangle (16.85, -0.05);
\node[font=\footnotesize\bfseries] at (14.4, -0.36) {Verifier / Auditor $V$ (Offline)};

\node[node box, minimum width=4.55cm, minimum height=1.45cm] (verifier_box) at (14.4, -1.52) {
  {\scriptsize \textbf{Audit Bundle:} Receipts $\{\mathsf{rcpt}_1,\dots,\mathsf{rcpt}_\ell\}$,}\\[-1pt]
  {\scriptsize signed reveal, and opened sample $\{d_i, s_i\}_{i\in\mathrm{Sel}}$}\\[2pt]
  {\scriptsize \textbf{Offline Verification:} Check signatures $(K, \{PK_{S_j}\}, PK_B)$,}\\[-1pt]
  {\scriptsize $t_{\mathrm{reg}} < T(a{+}2)$, round $a{+}2$ agreement, recompute $\mathrm{Sel}_\mu(c,\R_{a+2})$}
};

\draw[arrow] (1.75, 0.78) -- node[above, pos=0.45, font=\scriptsize\bfseries] {1. Parallel commit} node[below, pos=0.45, font=\scriptsize] {signed commitment, anchor $a$} (4.65, 0.78);

\draw[beacon arrow] (11.95, 1.15) -- node[above, font=\scriptsize\bfseries] {2. Beacon draw} node[below, font=\scriptsize] {$\R_{a+2}$ at $T(a{+}2)$} (9.15, 1.15);

\draw[arrow] (4.65, -0.78) -- node[above, font=\scriptsize\bfseries] {3. Signed receipts} node[below, font=\scriptsize] {$\{\mathsf{rcpt}_1,\dots,\mathsf{rcpt}_\ell\}$} (1.75, -0.78);

\draw[arrow, rounded corners=4pt] (0, -2.50) -- (0, -3.05) -- node[above, font=\scriptsize] {\textbf{4. Publish audit bundle:} receipts $\{\mathsf{rcpt}_j\}_{j=1}^\ell$ + signed reveal + opened sample $\{d_i, s_i\}_{i \in \mathrm{Sel}}$ \emph{(verified offline without contacting servers)}} (14.4, -3.05) -- (14.4, -2.50);

\end{tikzpicture}%
}
\caption{System architecture and decentralized trust model. The committer $C$ registers salted item hashes with $\ell$ independent servers before round $a+2$ of the decentralized randomness beacon $B$ exists. A signed anchor round $a$ locks all accepting servers to selection round $a+2$ without inter-server communication. Any auditor $V$ can later verify the receipts and opened sample offline; unbiased selection holds whenever the beacon threshold holds and at least one required server is honest.}
\label{fig:arch}
\end{figure*}
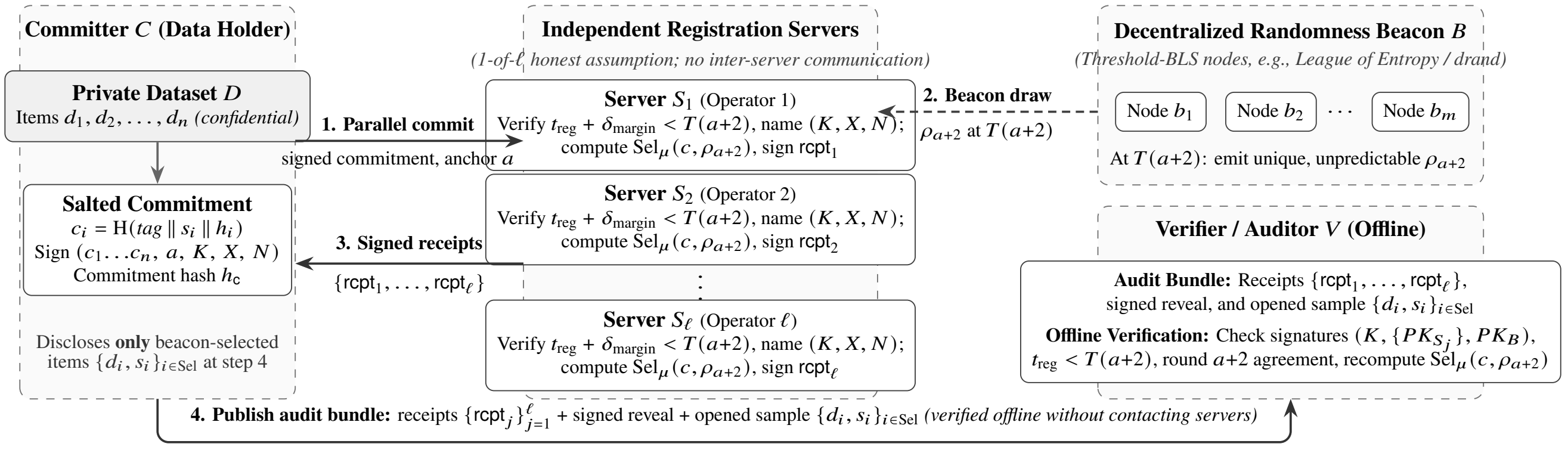

\section{Setting and Goals}
\label{sec:model}

\subsection{Use cases}

We use private AI benchmarks as the running example. Public test items
can contaminate training data, including through rephrasing that evades
n-gram checks \citep{yang2023rephrased,dekoninck2024}. Keeping a test set
private reduces direct exposure but leaves outsiders unable to inspect
its quality. A holder-selected sample does not resolve that problem.
The same sampling problem occurs in test fairness review, vulnerability
handling, compliance audits, and training-data provenance whenever the
population must remain private and the holder benefits from selective
disclosure.

\subsection{Parties}

\Cref{fig:arch} illustrates the system architecture and trust boundaries.
The committer $C$ holds private items $d_1,\dots,d_n$. \emph{Servers}
$S_1,\dots,S_\ell$ are run by independent operators; a server records when a
commitment arrives, obtains the beacon output, computes the selection, and
signs a receipt. It never sees the contents of unrevealed items. The \emph{beacon} $B$
publishes outputs $\R_1, \R_2, \dots$ on a public schedule, one per
\emph{round} $r$ at scheduled time $T(r)$, such that an output is
unpredictable before its scheduled time, publicly verifiable afterwards,
and unique (no party can produce a second valid output for a round). We
instantiate $B$ with the drand quicknet network of the League of Entropy
\citep{gailly2023tlock}, a threshold-BLS \citep{bls2001} beacon with a
3-second period \citep{drandquicknet}; any beacon with these properties can
replace it. A \emph{verifier} $V$ is anyone who holds the receipts and the
revealed items.

\subsection{Adversary and assumptions}
\label{sec:assumptions}

The committer is malicious: it wants the revealed sample to look better
than the committed set, and it may change data, time its submissions,
register many commitments under many keys, run its own server, and withhold
anything it has not published. Servers may be malicious individually; we
consider both a committer colluding with a server and a server acting
against a committer. The network is controlled by the adversary up to
denial of service.

\begin{enumerate}[label=A\arabic*.]
\item \label{asm:1} SHA-256 \citep{fips1804} is collision resistant and second-preimage
  resistant.
\item \label{asm:2} Ed25519 signatures \citep{ed25519} are existentially unforgeable.
\item \label{asm:3} Fewer than the threshold number of beacon nodes collude. The output
  of round $r$ is then unpredictable before $T(r)$ and, because threshold
  BLS signatures are unique, uniquely determined; the beacon itself cannot
  be ground. We model the output, the hash of the threshold signature, as
  uniformly random.
\item \label{asm:4} At least one server whose receipt the verifier accepts reports its
  registration time correctly from a clock within $\delta_{\mathrm{skew}}$
  of the beacon schedule (\cref{sec:anchor}).
\item \label{asm:5} For inputs fixed before a uniformly random key $\R$ is drawn, the
  values $\mathrm{HMAC\text{-}SHA256}_{\R}(\cdot)$ \citep{hmac} on those
  inputs are indistinguishable from uniform: HMAC is a pseudorandom
  function, and the committer chooses its inputs before the key exists.
\end{enumerate}
For temporal ordering, the verifier trusts a required server's
registration time; a valid signature authenticates that statement but
does not establish its truth. The name rule additionally requires servers
to enforce unique names and publish complete listings
(\cref{sec:grinding}). Signing keys are free to create in our model;
binding attempts to an accountable identity is a deployment requirement.
For privacy, we additionally model salted SHA-256 as a random oracle
with secret, unpredictable salts.

\subsection{Security goals}
\label{sec:goals}

Write $\mathrm{Sel}_\mu(c,\R)$ for the selection algorithm of
\cref{sec:selection} in mode $\mu$ applied to commitment $c$ and beacon
output $\R$. A \emph{bundle} is a receipt, a reveal, and revealed items. A
relying party fixes in advance a set of servers whose receipts it requires;
a verifier accepts a bundle only if every check of \cref{sec:verify} passes
for each of those receipts. Let $S^\ast$ be a required server that follows
the protocol (\asm{4}). The adversary controls the committer, all other servers,
and the network, and interacts with $S^\ast$ and the beacon over time. The
first three goals are statements about any accepted bundle whose
receipt from $S^\ast$ records selection round $r$ and mode $\mu$; the
fourth is a statement about all accepted bundles under one name.
\begin{description}[leftmargin=1em]
\item[Binding.] No efficient adversary can produce two different items
  $d \ne d'$ that the verifier accepts under the same committed hash, nor
  a commitment that the verifier attributes to a signing key whose holder
  did not sign it, nor a receipt attributed to $S^\ast$ that $S^\ast$ did
  not sign (\asm{1}, \asm{2}).
\item[Temporal ordering.] The item list of $c$ reached $S^\ast$, and was
  therefore fixed, before $T(r)$; by \asm{3} the output $\R_r$ was then
  unknown to every party.
\item[Unbiased selection.] The items the bundle labels as selected are
  exactly $\mathrm{Sel}_\mu(c,\R_r)$, and, over the beacon's coins for a
  $c$ fixed before $T(r)$, this set has the distribution stated in
  \cref{sec:selection} (\asm{5}).
\item[Name uniqueness.] If the relying party fixes a name $(K,X,N)$ and
  its required servers in advance, all accepted bundles under that name
  carry the same $\hc$, so the name yields at most one selection, and a
  relying party that asks the required servers after the selection round
  learns that $\hc$ whether or not the committer forwards its receipts.
\end{description}
The adversary breaks one of the first three goals if it obtains an
accepted bundle for which the statement is false, and the fourth if it
obtains two accepted bundles under the same name with different $\hc$. The first goal rests on the binding of the commitment and of
the signatures (assumptions \asm{1} and \asm{2}); the second on the registration
rule, the clock bound of assumption \asm{4}, and the beacon assumption \asm{3}; the
third follows from the first two and assumption \asm{5}; the fourth on
the name rule of \cref{sec:anchor} at $S^\ast$ and \asm{2}. Assumptions
\asm{1} to \asm{5} are those of \cref{sec:assumptions}.
\Cref{app:sketches} gives argument sketches, not proofs. Two things
are left to the deployment rather than stated as goals: repeated
commitments under other names or keys, which the deployment's identity
control limits (\cref{sec:grinding,sec:sybil}); and a required server
that misreports $t_{\mathrm{reg}}$ or its listings (\asm{4}).

\subsection{Scope}
\label{sec:nongoals}

The protocol establishes properties of the committed set; three things
are supplied by the process around it. \emph{Identity}: a commitment is
bound to a signing key, and the deploying protocol binds that key to an
accountable party, which is what limits repeated attempts
(\cref{sec:sybil}); the intended deployment of \cref{sec:lessons} admits one
invited node per institution. \emph{Reveal}: the deploying protocol
performs the reveal as a protocol step, and a verifier reports any
selected item without an opening as failed. \emph{Population}: sampling
the committed items shows their quality, not that nothing was left out,
a limit that audit standards state explicitly \citep{pcaobas2315} and
that election audits assign to a separate compliance audit
\citep{starkwagner2012}; an item committed as a hash of data not yet
written becomes visible when selected.

\section{Protocol}
\label{sec:protocol}

\begin{figure}[t]
\centering
\begin{tikzpicture}[font=\footnotesize,>=Stealth,x=0.9cm,y=0.72cm,
  box/.style={draw,rounded corners=2pt,minimum width=1.25cm,minimum height=0.45cm}]
  \node[box] (c) at (0,0) {Committer};
  \node[box] (s) at (2.9,0) {Servers};
  \node[box] (b) at (5.2,0) {Beacon};
  \node[box] (v) at (7.1,0) {Verifier};
  \foreach \n in {c,s,b,v} \draw[gray] (\n.south) -- ++(0,-5.0);
  \node[anchor=west,fill=white,inner sep=1pt] at (0.1,-0.65) {salt, hash, sign};
  \draw[->] (0,-1.45) -- node[above] {commitment, anchor $a$} (2.9,-1.45);
  \node[anchor=west,fill=white,inner sep=1pt] at (3.0,-1.85) {record $t_{\mathrm{reg}}$, name};
  \draw[->] (5.2,-2.65) -- node[above] {round $a{+}2$: $\R$} (2.9,-2.65);
  \node[anchor=west,fill=white,inner sep=1pt] at (3.0,-3.05) {select, sign};
  \draw[->] (2.9,-3.75) -- node[above] {receipt} (0,-3.75);
  \draw[->] (0,-4.55) -- node[above,fill=white,inner sep=1pt,yshift=2pt] {receipts, signed reveal with openings, items} (7.1,-4.55);
  \node[anchor=east,fill=white,inner sep=1pt] at (7.0,-4.95) {check offline};
\end{tikzpicture}
\caption{Protocol flow. The committer submits to all servers in parallel;
reveals can be published anywhere.}
\label{fig:flow}
\end{figure}
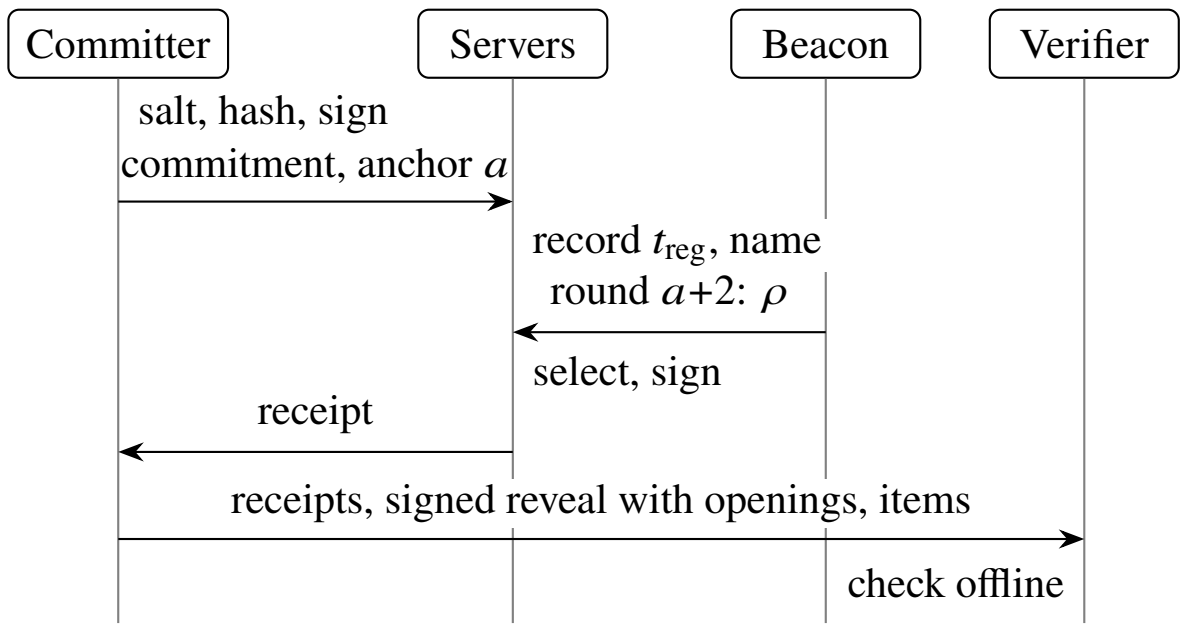

\Cref{fig:flow} shows the flow. The normative description is a versioned
specification with test vectors; it fixes byte-level encodings (canonical
serialization, domain-separation tags, fixed-width big-endian integers)
so that two independent implementations agree on every byte. This section
states object format 0.5.0 abstractly.

\subsection{Objects}
\label{sec:objects}

An item is a file or directory with item hash $h_i$, the SHA-256 of its
bytes or of a canonical manifest. The committer draws a 32-byte salt
$s_i$ per item and lists
\[
  c_i = \mathrm{SHA\text{-}256}(\mathit{tag} \,\|\, s_i \,\|\, h_i),
\]
which is binding under \asm{1} and hides $h_i$ under the secret-salt model.
The \emph{commitment} is the committer's signed canonical serialization
of the item commitments and their count $n$, the reveal probability
$\pi$ in parts per million, the mode $\mu$, the beacon reference with
anchor round $a$, an optional stream name $X$ with sequence number $N$,
metadata, its own timestamp, and its public key $K$; the SHA-256 of the
signed payload is the \emph{commitment hash} $\hc$, the same on every
server. A \emph{receipt} is a server's signed statement of the
commitment, its registration time $t_{\mathrm{reg}}$, and the selection
record (round, beacon output with proof, selected item commitments). A
\emph{reveal} is the committer's signed list of revealed items with one
\emph{opening} $(s_i, h_i)$ each. \Cref{app:objects} lists every field.

\subsection{Registration and selection round}
\label{sec:anchor}

The committer builds and signs the commitment and submits it to all
known servers in parallel. A server checks well-formedness and the
signature, rejects a duplicate, records $t_{\mathrm{reg}}$ from its own
clock, and applies two rules.

\paragraph{Anchor rule.}
The anchor round $a$ is the current round of the beacon schedule at
signing time. Every server selects with round
\[
  r = a + 2 ,
\]
and accepts the commitment only if, at its registration time,
\begin{align}
  T(a)   &\le t_{\mathrm{reg}} + \delta_{\mathrm{skew}}, \label{eq:anchor1}\\
  T(a+2) &>   t_{\mathrm{reg}} + \delta_{\mathrm{margin}}, \label{eq:anchor2}
\end{align}
with $\delta_{\mathrm{skew}} = \delta_{\mathrm{margin}} = 1$\,s.
Condition~\eqref{eq:anchor1} rejects anchors in the future, which would
only make the server wait arbitrarily long for the selection round.
Condition~\eqref{eq:anchor2} is the one that matters for security: the
selection round must not yet be available when the commitment is
registered. A server that receives the commitment too late must reject
it; it must not fall back to another round, and it must not select with
randomness that is already public. An unanchored rule that chooses the next round from each server's
registration time can instead split the same commitment across rounds
(\cref{sec:lessons}).

\paragraph{Proposition (agreement and temporal ordering).}
\label{prop:agreement}
Any two servers that follow the anchor rule and accept the same
commitment use round $a+2$, independently of their arrival times and
clocks. If a server also satisfies \asm{4} and
$\delta_{\mathrm{margin}} \ge \delta_{\mathrm{skew}}$, its registration
occurs before $T(a+2)$, while the output is unpredictable under \asm{3}. The argument is in \cref{app:proof}.

\paragraph{Name rule.}
A server registers at most one commitment per name $(K, X, N)$: a second
commitment with the same key, stream, and sequence but a different $\hc$
is refused, and the name stays taken after the commitment body itself has
left the server's retention window. The server publishes, signed, the
list of sequences it holds for $(K, X)$ and the list of streams it holds
for $K$, and each receipt states how many commitments the server holds
for the stream. \Cref{sec:grinding} analyzes what this gives.

\subsection{Selection}
\label{sec:selection}

The selected set is a deterministic function of the beacon output and
the commitment, so every server and verifier computes the same set. Let
$\R$ be the 32-byte randomness of the selection round. In
\emph{per-item mode}, each item receives a pseudorandom 64-bit number,
HMAC-SHA256 keyed by $\R$ over $\hc$ and the item commitment $c_i$, and is
selected if that number falls below a threshold proportional to the rate
$\pi$. Each distinct item is thus selected independently with probability
$p$ within $2^{-64}$ of $\pi/10^6$. In \emph{batch mode}, a seed computed
the same way from $\R$ and $\hc$ drives a partial Fisher-Yates shuffle
that selects exactly $m = \lceil \pi n / 10^6 \rceil$ items; rejection
sampling makes ideal draws yield a uniform $m$-subset. Binding the seed to
$\hc$ follows the $\mathrm{Hash}(S \,\|\, R)$ construction that NIST
describes for beacon-based sampling \citep{nistbeaconapps}.
\Cref{app:selection} gives the exact formulas.

\subsection{Verification}
\label{sec:verify}

A verifier authenticates the commitment, receipt, reveal, and beacon
output, checks the recorded registration time against the anchor
conditions, recomputes the selected set, and validates every opening
against the revealed data; a selected item without an opening fails the
audit. No step contacts a server, so a published bundle stays verifiable
after the servers disappear. A relying party that wants the name rule's
guarantee fixes its required servers in advance, demands a receipt
from each, and after the selection round asks them which commitment they
registered under the name (\cref{sec:grinding}). That query is the only
online step; it decides which audit took place, not whether a bundle is
valid. \Cref{app:verification} lists every check.

\section{Analysis}
\label{sec:security}

\subsection{Detection probability}
\label{sec:detection}

Suppose $k$ of $n$ committed items are \emph{bad}: fabricated, flawed, or
impossible to reveal. These are probabilities of selecting a bad item;
interpreting them as detection assumes review identifies its defect
or a missing opening is counted as failure. Per-item calculations assume
distinct selection inputs. In batch mode with $m = \lceil \pi n/10^6 \rceil$ selected
items, the probability that at least one bad item is selected is
\begin{equation}
  P_{\mathrm{batch}} = 1 - \binom{n-k}{m} \Big/ \binom{n}{m},
  \label{eq:batch}
\end{equation}
and in per-item mode it is
\begin{equation}
  P_{\mathrm{item}} = 1 - (1-p)^k .
  \label{eq:item}
\end{equation}

Detection depends mainly on $p$ and $k$, and on $n$ only when $k$ is a
sizable share of $n$ (\cref{tab:detection} in \cref{app:detection}). A random audit is
effective against widespread manipulation and weak against a handful of
bad items: at $p = 0.10$, fifty bad items are caught with probability above
0.99, a single one with probability 0.10. The two modes agree closely when
$k$ is small relative to $n$ but not in general (for $n=100$, $p=0.01$,
$k=50$: 0.500 in batch mode, 0.395 in per-item mode). The reveal
probability is a policy parameter that trades privacy of the committed set
against assurance.

A conditional calculation linking sampling rates to sanctions appears
in \cref{app:deterrence}.

\subsection{Attacks}

\Cref{app:attacks} summarizes defenses and their assumptions.
The following analysis addresses repeated attempts and residual trust.

\subsubsection{Repeated commitments and the name rule}
\label{sec:grinding}

A committer can create $g$ different commitments over the same data, for
example by changing the item order, the salts, or a dummy item, register
all of them, and publish only the receipt whose selection avoids the bad
items. Kroll et al.\ describe the analogous attack of drawing random
values repeatedly and keeping the preferred draw
\citep{kroll2017accountable}; the blockchain literature calls it
grinding \citep{kiayias2017ouroboros}. The anchor round does not prevent
it: all $g$ commitments can carry the same anchor. A deployment with
accountable identities sees the repeated submissions directly; the name
rule below gives a guarantee that does not depend on seeing them. In the ideal sampling model, distinct commitment-bound seeds give
independent selections, so a single attempt caught
with probability $P$ (\cref{eq:batch} or \cref{eq:item}) survives $g$
attempts with probability $1 - P^g$: at $n = 1{,}000$ and $p = 0.10$ in
batch mode, five hidden attempts let a committer with ten bad items
escape with probability 0.88 instead of 0.35, and with fifty bad items
0.022 instead of 0.0045 (per-item mode: 0.026 instead of 0.0052).

\paragraph{What the name rule gives.}
Append-only transparency logs address repeated attempts under a fixed
identity and epoch \citep{laurie2013rfc6962,rfc9162,tomescu2019}.
The name rule provides a narrower guarantee without a global log. A
relying party fixes $(K,X,N)$ and a required server set $\mathcal{S}$ in
advance, requiring a receipt from every server in $\mathcal{S}$ on the
same $\hc$. As long as at least one server $S^\ast \in \mathcal{S}$
honestly enforces name uniqueness and the anchor rule, $S^\ast$ signs at
most one commitment under $(K,X,N)$, so at most one commitment can obtain
all required receipts, even if the remaining $|\mathcal{S}|-1$ servers
collude with the committer. Splitting two commitments across honest
servers likewise leaves neither acceptable.

\paragraph{Abandoning a draw.}
The beacon output is public, so the committer knows the selection of its
commitment at $T(a+2)$, before any relying party sees a receipt. If a
registration reaches the relying party only through receipts the
committer forwards, the committer can abandon a commitment whose draw
looks bad, blame a server that did not respond, and retry under a new
name; the trials of \cref{sec:lessons} that did not complete at every
server make that excuse plausible. The relying party therefore does not
take the committer's word. After the selection round, and within the
servers' retention of commitment bodies, it fetches from every server in
$\mathcal{S}$ the signed listing for $(K,X)$ and, for an entry with
sequence $N$, the commitment, which must be signed by $K$ under that
name. Any such commitment binds the name: the audit is that commitment,
and without a complete bundle for it the audit fails. Two different
commitments under the name also fail it. A new name is acceptable only if
every server in $\mathcal{S}$ answers and none lists the old one; then no
commitment under the old name can complete, so none of its draws can be
kept. A commitment that can complete was registered at $S^\ast$ before
$T(a+2)$ and appears in this query whatever the committer forwards, and
an entry without a commitment signed by $K$ is ignored, so a server
cannot implicate a committer. The cost is that a genuine failure at one
server fails the audit instead of permitting a retry; selecting with a
later round than $a+2$ gives submissions more time to arrive. The
implementation provides this query as a library call and a command-line
tool.

Other names remain available. Signed stream and sequence listings can
expose extra registrations under $K$ if servers list them completely
and the relying party checks them. A party that accepts any name or key
is not protected. These are local registration rules, not a public
transparency log or a solution to Sybil resistance.

\Cref{app:trust} treats servers that misbehave, Sybil identities,
server timestamp trust, and what a commitment keeps private.

\section{Evaluation}
\label{sec:lessons}

The implementation shares a verification core across clients and
servers, and an independent implementation of hashing, item
commitments, selection, and the anchor rule agrees with it on
specification vectors and random inputs. For format 0.5.0 at
100{,}000 items, committing takes 200\,ms and offline receipt
verification 119\,ms (batch) or 239\,ms (per-item); commitments cost
67 bytes per item (\cref{app:cost}).

Round-selection experiments ran three local servers against the live
beacon with format 0.2.0 objects, whose anchor rule is that of
\cref{sec:anchor}; each trial submits one commitment to all three
servers, with the arrival at server $i$ delayed by $i\sigma/2$ for an
artificial spread $\sigma$ of 0 to 3\,s. Without an anchor, the three
servers split the same commitment across two rounds in a fraction
$\sigma/3$ of trials. With the anchor, across 630 trials (1{,}890
submissions; 420 of the trials deliberately delayed 1 to 6\,s after
anchoring), every one of the 807 accepted submissions used round $a+2$
and no two accepting servers ever disagreed. Anchoring to a public
relay's ``latest'' round, which often lags the schedule, had 70\% of
immediate submissions accepted (39 of 70 trials complete at all three
servers); anchoring to the later of the relay's round and the round
derived from the client's clock had 97\% accepted (126 of 140 trials
complete), losing trials only at spreads of 2.5 to 3\,s and never
reaching the client's retry path, which fires when no server accepts.
Tables and timing measurements are in \cref{app:eval}.

\begin{samepage}
\paragraph{Intended deployment: federated video benchmarking.}
A federated video benchmark in the PeerBench setting
\citep{cheng2025peerbench} integrates \sys{} to audit private test cases
held by independent institutions. Each institution runs one node under
one accountable identity, and benchmarks admit nodes by invitation,
which supplies the identity control of \cref{sec:sybil}. Institutions
commit to test cases and generated responses; the benchmark protocol
then reveals the beacon-selected cases and responses as a required step
and checks the \sys{} receipts, so disclosure is not left to the data
holder's discretion. Measurements from that deployment are future work.
\end{samepage}

\section{Related Work}
\label{sec:related}

\paragraph{Publicly seeded audit sampling.}
RFC~3797 selects committee members from a pool that is fixed and published
before pre-announced public random inputs exist
\citep{eastlake2004rfc3797}. The NIST beacon project describes audit
sampling in which a statement $S$ is published upfront and the seed is
$\mathrm{Hash}(S \,\|\, R)$ for a later beacon value $R$
\citep{nistbeaconapps}, and Rivest names that beacon as an alternative to
dice for seeding election audits \citep{rivest2018}. Risk-limiting
election audits draw a publicly seeded random sample from a ballot
manifest fixed before the draw \citep{lindeman2012rla}; batch mode is that
pattern with a beacon in place of dice. Kroll et al.\ combine
cryptographic commitments with fair random choices for the oversight of
automated decisions, discuss re-drawing, and propose a beacon seed
\citep{kroll2017accountable}. \sys{} follows this line. It adds
independent timestamping, receipts that verify offline, and an analysis of
multi-server deployment.

\paragraph{Evaluation integrity in machine learning.}
The PeerBench position paper argues for proctored, community-governed
evaluation and sketches the audit step that this paper builds
\citep{cheng2025peerbench}. TRUCE keeps test data private during
evaluation with trusted hardware or multi-party computation and audits
dataset quality by committing to each item and opening indices sampled by
the auditor \citep{rajore2024truce}. The randomness there is the auditor's
own, so third parties cannot re-check the selection. In concurrent work, Couteau et al.\ study interactive
proofs of proximity for model evaluation and derive reusable public audit
sets from a public random value such as a lottery or a beacon
\citep{couteau2026}; their setting is a proof system about a model, not a
commitment to private data with attested registration time, and they
report no implementation. Commit-reveal has also been applied to LLM-based
judging \citep{pardasani2026}, without beacon-selected reveals of private
data. Zero-knowledge approaches prove statements about evaluation or
training without revealing data or models
\citep{south2024zksnarks,waiwitlikhit2024}, at far higher cost and still
on top of a binding data commitment fixed before the audit.

\section{Limitations and Conclusion}
\label{sec:discussion}

\paragraph{Scope of the evidence.}
The guarantee covers the committed population; that it is the right
population is established by the surrounding process, as in election
audits \citep{lindeman2012rla}. Reveal, accountable identity, and the
required server set are supplied by the deploying protocol
(\cref{sec:nongoals}), and the deployment of \cref{sec:lessons} provides
all three. Registration timestamps rely on the named servers (\asm{4}). Our
experiments ran on a local testbed and did not measure wide-area delay
or clock drift. Revealed items may be sensitive, so deployments also
need data-handling rules.

\paragraph{Future work.}
A public transparency log of registrations; a signed pre-registration
object; a Merkle variant for batch mode; external timestamp anchoring;
soundness under a bounded number of repeated commitments; a deployment study with benchmark
maintainers.

\paragraph{Conclusion.}
\sys{} makes the selection of an audit sample independently verifiable
through signed commitments, registration attestations, and future public
randomness. The signed anchor gives agreement among accepting servers
without communication; experiments check that property and quantify the
availability cost of stale anchors. Receipts support offline verification
of a sample from the committed population. Their assurance remains
conditional on honest registration, controls on repeated attempts, and
external evidence that the population is the right one.

\paragraph{Availability.}
The reference implementation, specification, and test vectors are available
under the MIT license at \url{https://github.com/peerbench/nocherry}.

{\sloppy
\bibliographystyle{plainnat}
\bibliography{references}}

\appendix
\section{Protocol Details and Security Arguments}
\label{app:protocol}

\subsection{Argument sketches for the goals}
\label{app:sketches}

\paragraph{Binding.}
A revealed item is accepted only if it hashes to a committed item hash, and
the list of hashes is covered by the committer's signature and by $\hc$.
Substituting an item after commitment requires a second preimage (\asm{1}), and
forging a commitment under $K$ or a receipt under $S^\ast$ violates
signature unforgeability (\asm{2}). Changing the list yields a different $\hc$,
which is a different commitment with its own registration time and
selection.

\paragraph{Temporal ordering.}
\Cref{prop:agreement} places registration before the selection round
under \asm{3} and \asm{4}. The margin covers clock skew; delayed beacon publication
only postpones selection. A signed anchor number alone proves no signing
time, since the schedule is public. A malicious server can backdate its
attestation (\cref{sec:server-trust}).

\paragraph{Unbiased selection.}
Given a fixed commitment and an unpredictable beacon output, the selected
set is a deterministic function of values the committer did not know when
it committed. For distinct per-item inputs, \asm{5} gives computationally pseudorandom
draws with the effective probability $p$ of \cref{sec:selection}.
Batch mode uses rejection sampling to avoid modulo bias and yields a
computationally pseudorandom $m$-subset. The probabilities below use the
corresponding ideal sampling model.

\paragraph{Name uniqueness.}
Every accepted bundle under $(K,X,N)$ carries a receipt from $S^\ast$ on
its $\hc$, and $S^\ast$ cannot be impersonated (\asm{2}). By the name rule
$S^\ast$ signs receipts for at most one $\hc$ under $(K,X,N)$, so all such
bundles share that $\hc$ and, by temporal ordering and unbiased
selection, the same selection. Collusion among the other required
servers does not help, since every required receipt is demanded. A
commitment that can complete was registered at $S^\ast$, which lists it
under $(K,X,N)$ and serves it signed by $K$, so the relying party's query
of \cref{sec:grinding} finds it whatever the committer forwards.

\subsection{Objects}
\label{app:objects}
\paragraph{Item commitment.}
An item is a file or a directory with item hash $h_i$, the SHA-256 of
the file bytes or of a canonical manifest, which standard tools
reproduce. The committer draws a 32-byte salt $s_i$ per item (in the
reference client all salts are derived from one secret per commitment)
and lists
\[
  c_i = \mathrm{SHA\text{-}256}(\mathit{tag} \,\|\, s_i \,\|\, h_i).
\]
Under \asm{1} this is binding: two different openings imply a collision.
It hides the item hash under the separate random-oracle and secret-salt
model, rather than under collision resistance alone.

\paragraph{Commitment.}
A commitment contains the format version, the ordered list of item
commitments and their count $n$, the reveal probability as an integer
$\pi$ in parts per million, the selection mode $\mu$, a beacon reference
with the required anchor round $a$ (\cref{sec:anchor}), optionally a
stream name $X$ with a sequence number $N$, free-form metadata in
canonical form, the committer's own timestamp, its public key $K$, and
optionally an identity claim (\cref{sec:sybil}). The committer signs a
canonical serialization of all of these. The \emph{commitment hash} $\hc$
is the SHA-256 of the signed payload; it is the same on every server and
is the commitment's global identifier.

\paragraph{Receipt.}
A server-signed object that contains the commitment, the server's
registration time $t_{\mathrm{reg}}$, the latest beacon round at arrival,
the selection record (round, beacon output and its proof, mode, selected
item commitments), activity counts for the key and for the stream
(\cref{sec:grinding}), and the result of the identity check.

\paragraph{Reveal.}
A signed statement that lists $\hc$, the selected items being revealed,
any \emph{voluntary} items the committer reveals in addition (labeled
separately so that they cannot be mistaken for selected ones), and one
\emph{opening} $(s_i, h_i)$ per revealed item. The data itself can be
hosted anywhere; its locations are not part of the signed statement.

\subsection{Argument for the proposition}
\label{app:proof}
\emph{Argument.} Agreement follows because $a$ is signed and the rule
allows no fallback round. For temporal ordering, let $t$ be the actual
registration time. The clock bound gives
$t \le t_{\mathrm{reg}}+\delta_{\mathrm{skew}}
\le t_{\mathrm{reg}}+\delta_{\mathrm{margin}} < T(a+2)$.
Agreement is thus a construction property; freshness additionally needs
the clock and beacon assumptions. Neither guarantees that all required
servers accept: a late arrival is rejected, and an adversarial network
can prevent completion.

With a 3\,s beacon period and 1\,s margins, a client that anchors to its
local clock has 2 to 5\,s from signing to reach every server
(\cref{app:beacon}).

\subsection{Selection details}
\label{app:selection}
Let $\R$ be the 32-byte randomness of the selection round. The mode is
the one the committer signed; a receipt in another mode fails
verification. All arithmetic is on integers.

\paragraph{Per-item mode.}
For distinct item commitments, per-item selection approximates
independent draws with target probability $\pi/10^6$:
\begin{align*}
  v_i &= \mathrm{be64}\bigl(\mathrm{HMAC\text{-}SHA256}_{\R}(\mathit{tag} \,\|\, \hc \,\|\, c_i)\bigr),\\
  \text{select } i &\iff v_i < \lfloor \pi \cdot 2^{64} / 10^6 \rfloor,
\end{align*}
where $\mathrm{be64}$ reads the first eight bytes as a big-endian integer.
At $\pi = 10^6$ the threshold is $2^{64}$ and every item is selected. In the ideal random-function model the effective probability is
$p=\lfloor\pi 2^{64}/10^6\rfloor/2^{64}$, differing from the target by
less than $2^{-64}$. Repeated $c_i$ produce the same draw, so the
independence guarantee requires distinct inputs. The number selected is
random, which suits a committer who registers items one at a time (such a stream has a detection guarantee only once it is named,
\cref{sec:grinding}).

\paragraph{Batch mode.}
Exactly $m = \lceil \pi n / 10^6 \rceil$ items are selected by a partial
Fisher-Yates shuffle of the committed order, driven by draws
$\mathrm{be64}(\mathrm{HMAC\text{-}SHA256}_{\sigma}(\mathit{counter}))$
from the seed
\[
  \sigma = \mathrm{HMAC\text{-}SHA256}_{\R}(\mathit{tag} \,\|\, \hc),
\]
with rejection sampling in place of modulo reduction, so ideal uniform draws yield a uniform $m$-subset without modulo bias.
With HMAC the guarantee is computational, under the pseudorandom-function
assumption. Binding the seed to $\hc$ follows the
$\mathrm{Hash}(S \,\|\, R)$ construction that NIST describes for
beacon-based sampling \citep{nistbeaconapps}.

\subsection{Complete verification checklist}
\label{app:verification}
Given a receipt, a reveal, and the revealed data, a verifier checks:
\begin{enumerate}
\item that the commitment is well formed and the committer's signature
  and the recomputed $\hc$ hold;
\item the server's signature on the receipt;
\item that the beacon output is authentic for the stated beacon and round
  (for drand, the BLS signature under the chain's public key and
  randomness equal to the hash of the signature);
\item that the selection round equals $a+2$ and that $t_{\mathrm{reg}}$
  precedes $T(a+2)$ (the verifier applies neither the margin nor
  condition~\eqref{eq:anchor1}, which guard the server's own clock; a
  server that follows the rule never issues a receipt inside the margin);
\item that the selection record uses the signed mode and probability and
  that recomputing the selection from the commitment and the beacon
  output yields exactly the selected items;
\item the signature on the reveal;
\item that the reveal names this commitment, that every revealed item is
  a committed item, that only items the beacon selected are listed as
  selected, and that every opening recomputes its item commitment;
\item that each revealed item's data hashes to the opened $h_i$;
\item completeness: every selected item is revealed. A selected item
  without a reveal is reported and the result is not ``verified''; a
  relying party should count such an item as failed, not as a missing
  data point.
\end{enumerate}

All checks are required for acceptance. A partial result does not establish
the security goals of \cref{sec:goals}.

\subsection{Commitment and verification cost}
\label{app:cost}
Microbenchmarks measure format 0.5.0 under Node.js 24 on an Apple M3 Max
laptop. The round-selection experiments of \cref{sec:lessons} use format
0.2.0 objects, whose anchored round and acceptance conditions are
identical; they check the anchor implementation, not all security
properties of 0.5.0, which has not been deployed in production.

At 100{,}000 items,
salting and committing takes 200\,ms and signing 85\,ms; offline receipt
verification (including beacon signature and selection recomputation)
takes 119\,ms in batch mode and 239\,ms in per-item mode; verifying the
reveal of 10{,}000 selected items takes 316\,ms and 454\,ms. Commitments
cost about 67 bytes per item and reveals 291 bytes per revealed item.
Hashing runs at 400\,MiB/s (25\,s for 10\,GiB). The scaling bottleneck is
commitment size, motivating a Merkle variant for batch mode.

\section{Attacks and Residual Trust}
\label{app:trust}

\subsection{Servers and identities}
\paragraph{Servers.}
A server that drops entries from its listing, registers a second
commitment under a taken name, or backdates $t_{\mathrm{reg}}$ is not
detected by any check on a single receipt; the defense is the relying
party's required set $\mathcal{S}$, where all servers in $\mathcal{S}$
would have to collude to forge a full bundle under a fixed name. A public
transparency log of all registrations would make omissions detectable; the
protocol does not include such a log.

\paragraph{Sybil resistance.}
\label{sec:sybil}
Free signing keys allow a fresh identity for each attempt. A signed DID
claim lets a server attest whether the key belongs to that identifier,
but does not make identifiers scarce or link a holder's other identities.
The control on attempts therefore comes from the deployment: in the
intended deployment of \cref{sec:lessons}, each institution operates one
invited node under one identity, so a committer cannot multiply attempts
across keys.

\subsection{Server timestamp trust}
\label{sec:server-trust}

A malicious server that colludes with the committer can register a
commitment after the selection randomness is public and assert an earlier
registration time. A verifier cannot detect this from a single receipt.
The anchor does not help here: it fixes which round is used, not when the
commitment was made. The defense is again the relying party's required
server set $\mathcal{S}$, all of which would have to collude to backdate
a complete bundle. Assumption \asm{4} ($1$-of-$|\mathcal{S}|$ honest timestamp
reporting) is therefore close to the property one would like to prove, and
we state it as an assumption for that reason. Two strengthenings are future
work: signing the beacon output of round $a$, not only its number, would
prove signing after $T(a)$ without trusting any server, at the price of a
shorter window; an external timestamping service
\citep{rfc3161,opentimestamps} would prove existence before $T(a+2)$.

\subsection{What stays private, and refusal}

Before reveal a commitment discloses $n$, $\pi$, $\mu$, the stream name
and sequence, and the metadata. Item hashes are hidden under the
secret-salt model; metadata can still disclose information about items. Repeated audits
of one private set still disclose a growing share of it; the
reusable-holdout literature \citep{dwork2015holdout} offers a way to
reason about how many rounds a private benchmark can sustain. A committer
that refuses to reveal a selected item, or committed to an item with no
data behind it, produces a selected item without an opening; the verifier
reports it and relying parties should count it as failed, as audit
standards do for a selected item to which no procedure can be applied
\citep[paragraphs .25 and .40]{pcaobas2315}.

\subsection{Attack summary}
\label{app:attacks}

\begin{table}[H]
\centering
\small
\begin{tabular}{@{}>{\raggedright\arraybackslash}p{0.46\columnwidth}>{\raggedright\arraybackslash}p{0.48\columnwidth}@{}}
\toprule
Attack & Defense or boundary \\
\midrule
Altered items or signed fields & Hash binding and signatures \\
Biased or mislabeled selection & Signed mode, bound seed, recomputation \\
Split selection rounds & Anchor; late servers reject \\
Commit after the draw & Honest timestamp and clock bound (\asm{4}) \\
Guessed-item membership tests & Salted commitments; hiding model \\
Repeated audit attempts & Name rule; other names and keys remain a risk \\
Abandoning a drawn commitment & Relying party queries $\mathcal{S}$; a bound name without a bundle fails \\
Backdating or omitted listings & Trust in required servers ($\mathcal{S}$) \\
Missing selected items & Incomplete audit; count as failure \\
Omitted population items & Requires external evidence \\
\bottomrule
\end{tabular}
\caption{Defenses and assurance boundaries in format 0.5.0.}
\label{tab:attacks}
\end{table}

\subsection{Deterrence under an assumed sanction}
\label{sec:deterrence}
\label{app:deterrence}

A small sample can deter cheating under an incentive model
\citep{becker1968}; detection certainty also matters empirically
\citep{nagin2013}. If a risk-neutral committer gains $G$ when one bad item
escapes detection and loses $R$ when caught, honesty is at least as good
as cheating when $p \ge G / (G + R)$. At $p=0.10$ or $0.05$, the sanction
must be at least $9\times$ or $19\times$ the gain. With $g$ hidden
independent attempts, detection falls to $p^g$; at $p=0.10$, five attempts
require $R/G \ge 99{,}999$. Audit rates therefore require both controls on
repeated attempts and credible sanctions.

\subsection{Detection probabilities}
\label{app:detection}
\Cref{tab:detection} tabulates \cref{eq:batch} for representative $n$, $p$, and $k$.

\begin{table}[H]
\centering
\small
\begin{tabular}{@{}rrrcccc@{}}
\toprule
$n$ & $p$ & $m$ & $k{=}1$ & $k{=}5$ & $k{=}10$ & $k{=}50$ \\
\midrule
100    & 0.01 & 1    & 0.010 & 0.050 & 0.100 & 0.500 \\
100    & 0.05 & 5    & 0.050 & 0.230 & 0.416 & 0.972 \\
100    & 0.10 & 10   & 0.100 & 0.416 & 0.670 & 0.999 \\
1{,}000  & 0.10 & 100  & 0.100 & 0.410 & 0.653 & 0.996 \\
10{,}000 & 0.10 & 1{,}000 & 0.100 & 0.410 & 0.651 & 0.995 \\
\bottomrule
\end{tabular}
\caption{Probability that at least one of $k$ bad items is selected in
batch mode (\cref{eq:batch}).}
\label{tab:detection}
\end{table}

\section{Evaluation Details}
\label{app:eval}

\begin{table}[H]
\centering
\small
\setlength{\tabcolsep}{4pt}
\begin{tabular}{@{}rrrrr@{}}
\toprule
 & Unanchored & \multicolumn{3}{c}{Anchored} \\
\cmidrule(lr){3-5}
$\sigma$ (s) & disagree & \multicolumn{1}{c}{relay accepted} & \multicolumn{1}{c}{clock accepted} & disagree \\
\midrule
0   & 0/60  & 24/30 & 60/60 & 0 \\
0.5 & 7/60  & 24/30 & 60/60 & 0 \\
1.0 & 19/60 & 15/30 & 60/60 & 0 \\
1.5 & 30/60 & 18/30 & 60/60 & 0 \\
2.0 & 42/60 & 27/30 & 60/60 & 0 \\
2.5 & 53/60 & 20/30 & 56/60 & 0 \\
3.0 & 60/60 & 19/30 & 50/60 & 0 \\
\bottomrule
\end{tabular}
\caption{Round agreement and acceptance with three local servers and a
live beacon. Unanchored disagreement is per trial (60 trials per spread).
Anchored acceptance is per submission: 10 relay-anchor trials and 20
clock-anchor trials per spread, each contacting three servers. The final
column counts trials (of 30 per spread) with disagreement among accepting
servers; zero disagreement does not imply acceptance by all required
servers. Relay-anchor acceptance varies with relay staleness rather than
with spread.}
\label{tab:multi}
\end{table}

Round-selection experiments use three local servers and the live
beacon. The local testbed isolates arrival spread; it does not measure
wide-area delay or independent clock drift. Raw data and scripts
accompany the paper source.

\subsection{Round agreement}
\label{app:agreement}
An unanchored rule can select different rounds when the same commitment
reaches servers on opposite sides of a beacon tick. To quantify this
and check the anchor rule, we submitted
fresh commitments over 20 items with $p = 0.1$, with and without an anchor, delaying the submission to server $i$ by $i\,\sigma/2$ for a spread $\sigma$ from 0 to
3\,s (\cref{tab:multi}). Without an anchor the disagreement rate is
$\sigma/3$ within sampling error, as expected when the first registration
falls uniformly in the period; in 208 of the 211 split trials the two
rounds also produced different selections. The anchored runs comprise 210 immediate trials (70 relay-anchor and
140 clock-anchor; their 630 submissions appear in \cref{tab:multi}) plus
420 trials delayed 1 to 6\,s after choosing a relay anchor, 1{,}890
submissions in all. Every one of the 807 accepted submissions used round $a+2$, no two accepting
servers ever disagreed, and in every submission the server's decision
matched the rule of \cref{sec:anchor} applied to its recorded time. This
is an empirical check of the implementation against
\cref{prop:agreement}, not a proof of it.

\subsection{Anchor freshness and availability}
Fresh anchors improve acceptance. A client that anchors to the round a
public relay reports as ``latest'' had only 70\% of its immediate
submissions accepted (147 of 210 across all spreads, and 39 of 70 complete
three-server trials), and none when it waited 4\,s; every rejection cited
an already-available selection round caused by relay pointer lag. A client
that instead anchors to the later of the relay's round and the round
derived from its own clock had 97\% of submissions accepted (406 of 420,
and 126 of 140 complete three-server trials), never reached the
client's retry path (which fires only when no server accepts), and lost
only submissions whose artificial inter-server spread reached 2.5 to
3\,s. Relay pointer staleness and receipt latency measurements appear in
\cref{app:beacon}.

\subsection{Beacon, relay, and anchor timing}
\label{app:beacon}
With a 3\,s period and $\delta_{\mathrm{skew}} = \delta_{\mathrm{margin}} = 1$\,s,
conditions~\eqref{eq:anchor1}--\eqref{eq:anchor2} leave a 6\,s window
$[T(a) - 1\text{\,s},\, T(a+2) - 1\text{\,s})$. A client anchoring at
$T(a) + x$ ($0 \le x < 3$\,s) has $5 - x$ seconds (2 to 5\,s) to reach
every server; if none accepts, the client signs a new commitment with a
fresh anchor and retries once.

In \cref{sec:lessons}, the relay's ``latest'' pointer named the current
round in 297 of 490 fetches, a round 3 to 5\,s old in 136, and a round
at least 5\,s old, for which condition~\eqref{eq:anchor2} can no longer
hold at any server, in 57. Taking the later of the relay's round and the clock-derived
round avoids stale anchors at the cost of waiting for round $a+2$:
receipts arrived after 2.1\,s (median) and 3.2\,s (p90) unanchored,
3.4\,s and 5.3\,s with relay anchors, and 4.2\,s and 6.0\,s with clock
anchors. Polling each upcoming round by number every 100\,ms at two public
relays over 240 rounds, a round was first served 0.16\,s (median) and
1.1\,s (max) after schedule at one relay, and 0.68\,s and 1.65\,s at the
other, never early; the ``latest'' pointer trailed by 1.2\,s (median) and
up to 5.3\,s. Condition~\eqref{eq:anchor2} relies only on rounds never
appearing early; the margin covers clock skew, not relay latency.

\subsection{Other randomness sources}
\label{app:beacons}
A commitment names its beacon by type and chain hash, and the verifier
recomputes the selection from the beacon output recorded in the receipt,
so the beacon is a replaceable component; the reference implementation
accepts only drand quicknet. A replacement must satisfy \asm{3}: outputs
unique, unpredictable before a scheduled time, and publicly verifiable
afterwards. Candidate designs include threshold-BLS beacons such as
DFINITY's \citep{hanke2018dfinity}, the VRF-based NEAR beacon
\citep{skidanov2019near}, and the VRF-derived block seed of Algorand
\citep{gilad2017algorand}; each needs its own output verification and an
assessment against \asm{3} before use. Ethereum's RANDAO does not qualify,
since proposers of the last slots in an epoch can bias it
\citep{alpturer2024randao}. The anchor rule also needs a known schedule
$T(r)$; for a beacon whose rounds follow block production, the rule
would be restated in block heights with a margin for block-time
variance.

\paragraph{Related beacon work.}
drand descends from work on scalable bias-resistant distributed randomness
\citep{syta2017}. Surveys cover the design space, including beacons based
on verifiable delay functions and commit-reveal beacons with their
last-revealer problem \citep{choi2023sok,kavousi2023sok}. We use
commit-reveal to audit data, not to generate randomness, and treat the
beacon as an interchangeable component. The closest application of drand
that we know is the CURBy beacon, whose inputs are committed before a
drand pulse seeds a randomness extractor \citep{kavuri2024curby}.

\end{document}